# Numerical Renormalization Group Calculations for Similarity Solutions and Travelling Waves

LIN-YUAN CHEN* AND NIGEL GOLDENFELD

*Department of Physics and Materials Research Laboratory,*

*University of Illinois at Urbana-Champaign, 1110*

*West Green Street, Urbana, Illinois 61801-3080, USA.*

## ABSTRACT

We present a numerical implementation of the renormalization group (RG) for partial differential equations, constructing similarity solutions and travelling waves. We show that for a large class of well-localized initial conditions, successive iterations of an appropriately defined discrete RG transformation in space and time will drive the system towards a fixed point. This corresponds to a scale-invariant solution, such as a similarity or travelling-wave solution, which governs the long-time asymptotic behavior. We demonstrate that the numerical RG method is computationally very efficient.

---

* Present address: Institute for Theoretical Physics, University of California, Santa Barbara, CA 93106.

## 1. Introduction

Recently, renormalization group (RG) theory has been applied to a number of non-equilibrium physical systems[1−10] without any statistical aspect, to study the long-time or large-scale intermediate asymptotic behavior. In particular, the important cases of similarity solutions[1−9] $u(x,t) = t^{-\alpha} f\left(xt^{-\beta}\right)$, or travelling-wave solutions[10] $u(x,t) = f\left(x - vt\right)$ were treated. In all of these cases, generically referred to as exhibiting intermediate asymptotics of the second kind,[1] the exponents $\alpha, \beta$ or the velocity $v$ cannot be determined by simple dimensional analysis or from conservation laws, but are determined only by solving the full problem itself. Previous work demonstrated that these exponents are the *anomalous dimensions* of the field theoretic RG, and may be calculated using RG[2,3,11,12] to remove systematically divergent or secular terms from a naive perturbation expansion.

Although RG is often formulated in connection with perturbation theory, it is essentially *non-perturbative* and has a direct geometrical interpretation, as Wilson showed in the context of critical phenomena.[13] The purpose of this present paper is to explore the long-time asymptotic behavior of certain simple non-equilibrium systems within this geometrical picture. We will see that the fixed points of the appropriately defined RG transformation correspond to scale-invariant solutions, such as similarity or travelling-wave solutions. The simple numerical method given here, which exploits this picture, extends the practical applicability of the RG to problems where there is no small perturbation parameter, in contrast to our previous perturbative work.

Other numerical methods, based on rescaling and thus closely related to the RG, have been applied to study (*e.g.*) finite time blow-up in the two-dimensional nonlinear Schrodinger equation[14–16] and in axisymmetric three dimensional Euler dynamics.[17] Numerical RG methods, particularly Monte Carlo RG, are widely used in equilibrium statistical mechanics to calculate equilibrium phase diagrams and anomalous dimensions.[6] Monte Carlo RG has also been applied to a particular non-equilibrium system — the driven diffusive gas — to calculate scaling exponents of the nonequilibrium steady states.[18]

We begin by considering numerical RG transformations to obtain self-similarity solutions. A RG transformation $R_{b,\beta}$, which depends on two parameters, a dilation parameter $b$ and an exponent $\beta$, is defined on the space of functions $u(x,t)$ at some arbitrary time $t$ by $u'(x,t) = R_{b,\beta}\{u(x,t)\}$. This transformation involves the following three steps:

(1) Evolve the function $u(x,t)$ forward over a *finite* time, from $t$ to $t' = bt$, $b > 1$, using the governing PDE, and call the result $\hat{u}(x,t')$. This step is analogous to the block spin transformation used in critical phenomena.

(2) Rescale $x$ by defining $x' = b^{-\beta}x$, so that $\tilde{u}(x',t') = \hat{u}(x'b^{\beta},t')$.

(3) Rescale the function $u$ itself by an amount $Z(b) \equiv \hat{u}(0,t')/u(0,t)$, so that $u'(x',t') = Z(b)\tilde{u}(x',t')$.

Thus, the resulting RG transformation yields $u'(x,t) = Z(b)u\left(b^{\beta}x, bt\right)$. The basic idea is that any fixed point $u^* = R\{u^*\}$ corresponds to a similarity solution. For the simple systems discussed in this paper, only one fixed point exists, which is a stable attractor. We will see that repeated iterations of the RG transformation drive the function $u$ to its fixed point value more rapidly than straightforward time evolution for large times. The semi-group property of this RG transformation $(b > 1)$ : $R_{b_1,\beta}R_{b_2,\beta} = R_{b_1 b_2,\beta}$, simply implies $Z(b) = b^{\alpha}$ or equivalently $\alpha = d\log Z/d\log b$. The parameter $\beta$ must be varied to search for fixed points of this RG transformation, which may not exist for arbitrary $\beta$. For appropriate values of $\beta$, after an infinite number of successive iterations of the RG transformation defined above, the fixed point, $u^*(x,t)$, will be attained, if the initial conditions are in the basin of attraction of the fixed point, *i.e.*, we have $u^*(x,t) = b^{n\alpha} u\left(b^{n\beta}x, b^n t\right)$, as $n \to \infty$. By setting $b^n = A/t$, we find for any constant $A$, $u^*(x,t) = t^{-\alpha}\phi\left(xt^{-\beta}\right)$, as $t \to \infty$, where $\phi$ is a scaling function to be determined. In principle, the RG transformations outlined above – either continuous or discrete – can be implemented in many different ways, but here we will use a discrete formulation, which is more useful for numerical applications.

## 2. Numerical Method for Similarity Solutions

In our numerical implementation, we adopt a finite-difference Euler scheme to discretize the continuous version of the PDE at hand. We write $t = i\Delta t, x = j\Delta x, i = 0,1,2,\cdots,M, j = 0,1,2,\cdots,N$, with $\Delta t$ a time step and $\Delta x$ a mesh size. By simply evolving the PDE over a short, finite time from $t = t_0$ to $t_1 = bt_0$ with $n_0 = (b-1)t_0/\Delta t$ time steps, we obtain a set of data $\left\{u'(x = j\Delta x, t_1 = bt_0)\right\}$. The step (2) is best realized by rescaling *the mesh size* $\Delta x$, rather than by changing the discrete sites $j = 0,1,2,\cdots,N$ at each iteration. Thus, after one iteration, the new mesh size is $(\Delta x)_1 = b^{-\beta}\Delta x$. Finally, we implement step (3) by rescaling the whole discrete set of data by a factor $Z_1(b) = u'(0, bt_0)/u(0,t_0)$. Thus, after the first RG iteration, a set of coarse-grained data is obtained at fixed sites $j$:

$$u^{(1)}(x_1, t_0) = Z_1(b)u'(b^\beta x_1, bt_0), \tag{2.1}$$

with $x_1 = j(\Delta x)_1, j = 0,1,2,\cdots$. This data set will constitute the new initial condition for the second RG iteration. The time step $\Delta t$ and the total number of time steps $n_0 = (b-1)t_0/\Delta t$ are kept fixed through all the RG iterations, so that after $n$ RG iterations, we will have calculated the $n$-th iteration

$$u^{(n)}(x_n, t_0) = Z_n(b)\cdots Z_1(b)u'\left(b^{n\beta}x_n, b^n t_0\right), \tag{2.2}$$

where $x_n = j(\Delta x)_n$ with a rescaled mesh size $(\Delta x)_n = b^{-n\beta}\Delta x$. Thus, in our numerical RG transformation, the equation is always solved over a finite time interval, and the system size $N(\Delta x)_n$ continues to shrink with the number of RG iterations increasing. This behavior can be understood by noting that the limit $x/t^\beta \to 0$ in a possible similarity solution $u(x,t) = t^{-\alpha}\phi(x/t^\beta)$ can be realized by letting $x$ (or mesh size $\Delta x$) $\to 0$ and keeping $t$ fixed at a finite value, instead of letting $t \to \infty$ and keeping $x$ fixed. Therefore, the numerical RG method is conceptually different from direct numerical integration (DNI), in which we simply evolve the PDE over a very long time $t$ and generally have to choose a very large-size system.

In practice, a potential difficulty is the choice of parameter $\beta$. For the fixed point, $\beta$ is determined by requiring that, if the curve $t^\alpha u(x,t)$ vs. $xt^{-\beta}$ is plotted, all the data should collapse onto a single curve, namely the universal scaling function $\phi$. A simple alternative is to require that, if $\log u(x=0,t)$ vs. $\log t$ is plotted, all the data should lie on a straight line, enabling one to read off the slope $-\alpha$. In practice, the correct value $\beta$ is selected in the following way. Given an estimated interval for $\beta$, $\beta_1 < \beta < \beta_2$, we calculate in the last RG iteration, the slopes $d\log u(x=0,t)/d\log t$ at $t = t_0$, $\alpha_1$, and at $t = bt_0$, $\alpha_2$, respectively. For arbitrary values of $\beta$, the difference between $\alpha_1$ and $\alpha_2$, should be a function of $\beta$, *i.e.* $\Delta\alpha = \alpha_2 - \alpha_1 = f(\beta)$. Only for certain $\beta$, will $\Delta\alpha$ be equal to zero. This is a typical root finding problem, if we regard $\beta$ as the root of the function $f$. The appropriate value of $\beta = \beta^*$ is then easily found using any given root finding algorithm. We have found this method to converge rapidly, and still be faster than DNI.

The first example we consider is the so-called modified porous-medium equation

$$\partial_t u = D\Delta u^{1+n}, \tag{2.3}$$

where $\Delta$ represents a $d$-dimensional Laplacian, $n$ is arbitrary, $D = 1$ for $\partial_t u \geq 0$ and $D = 1 + \epsilon$ for $\partial_t u \leq 0$. This problem has previously[5] been treated analytically using the RG method, for $\epsilon \ll 1$. We have applied the numerical discrete RG transformation to several special cases, including the problem of gravity-driven groundwater $(n = 1, d = 2)$. For simplicity, we will take the Barenblatt equation $(n = 0, d = 1)$ to illustrate our basic ideas on numerical RG. This equation describes the pressure $u$ during filtration of an elastic fluid in an elasto-plastic porous medium. For $\epsilon \neq 0$, the material exhibits hysteresis and the diffusion decays anomalously, with the long-time asymptotics of the similarity form $u(x,t) \sim t^{-(1/2+\alpha)} f(xt^{-1/2}, \epsilon)$, where the anomalous dimension is calculated using perturbative RG[6] to be $\alpha = \epsilon/(2\pi e)^{1/2} + O(\epsilon^2)$.

A simple, *explicit*, discrete scheme can be written as

$$u(j, i+1) = u(j,i) + Dr\Delta u(j,i), \tag{2.4}$$

where $u(j,i) = u(x = j\Delta x, t = i\Delta t), r = \Delta t/(\Delta x)^2$, the discrete Laplacian is

$$\Delta u(j,i) = u(j+1,i) - 2u(j,i) + u(j-1,i), \tag{2.5}$$

and $D = 1$ if $\Delta u(j,i) \geq 0$, $D = 1 + \epsilon$ if $\Delta u(j,i) < 0$. The condition for numerical stability $Dr \leq 1/2$ must be imposed in this explicit scheme. The numerical values of the parameters $b, \Delta t, \Delta x$ are chosen in such a way as to not only result in the quickest rate of convergence to the true solution, but also to attain sufficient accuracy. Typical values of $\Delta t$ and $\Delta x$ are chosen as 0.01 and 2.0, respectively, because a too small value for $r = \Delta t/(\Delta x)^2$ gives a high accuracy, but needs too large a number of time steps $n_0$, whilst a too large value for $r$ (which still satisfies the numerical stability condition) is not accurate enough. Although in principle any value $b > 1$ can be chosen, we choose a dilation parameter $b$ typically about 1.02. For too large a value of $b$, large number of time steps $n_0$ would be required, and the system would shrink too quickly even after only a few number of RG iterations (the mesh size at $n$-th round is $(\Delta x)_n = \Delta x/b^{n\beta}$). For a very small $b$, a large number of RG iterations are necessary to produce good enough accuracy, and this takes too much computer time. Thus, we must adopt a compromise between the number of RG iterations and the size of the finite time interval used in each iteration. The typical number $N$ of RG iterations is chosen as $100 \leq N \leq 500$, depending upon how accurate the solution is required to be, whilst satisfying the condition for numerical stability $D\Delta t/(\Delta x)_N^2 \leq 1/2$ even after $N$ RG iterations.

In figure 1, we show how the anomalous exponent $\alpha$ extracted from $Z(b)$ varies with the number of RG iterations $N$ and converges to the value $\alpha = 0.6975$ in the case $\epsilon = 1.0$, starting from different localized initial conditions, where we have taken only a small number of time steps $n_0$. As we see, an advantage of our numerical RG is that, even if we choose a very small number of time steps $n_0$ (*e.g.* $n_0 = 10$) in each RG iteration (an obviously poor approximation), after a sufficient large number of RG iterations, the final result for the exponent $\alpha$ is still very accurate.

For purpose of comparison, we also perform a conventional DNI of the Barenblatt equation in $d = 1$ and 2 dimensions using the same explicit scheme. We choose the same initial conditions as in the numerical RG calculation, $\Delta x = 2.0$ and $\Delta t$ as large as possible, whilst satisfying the numerical stability condition $r = \Delta t/(\Delta x)^2 \leq 1/2d$. We compare both CPU times consumed on the same computer and the total number of iterations performed so that the same accuracy of the solution (*e.g.* the same relative error) is reached.

Suppose that the number of time steps taken to have the same accuracy is $n_1$ and $n_2$, respectively, in the DNI and the numerical RG method, and the number of RG iterations is $N$. An initial condition with compact support will be non-zero at most over $n+1$ additional spatial grid points after the $n$-th time iteration. Therefore, the total number of iterations is estimated to be of order $N_1 = O(n_1^{d+1})$ and $N_2 = O(N n_2^{d+1})$ for $n_1, n_2 \gg 1$, respectively, in $d$ dimensions. If $n_1/n_2 \gg 1$, then $N_1 \gg N_2$, which implies that DNI is much slower than the numerical RG method. To achieve this goal, we choose $N$ typically about $100 - 500$, and $n_2$, typically about $20 - 50$.

Depending upon the initial conditions, the explicit numerical RG scheme is about $5 - 10$ times faster than DNI in $d = 1$ dimension, and about $50 - 100$ times in $d = 2$ dimensions, where the requisite accuracy is $1 - 2\%$. We find that, in DNI, the more unsymmetrical and unsmooth the initial conditions are, the more slowly the solutions converge, especially when a very high accuracy is required. It is not unreasonable to expect that our numerical RG method will be about several hundred times faster than DNI in $d = 3$ dimensions, with arbitrary, unsymmetrical, localized initial conditions. Therefore, it seems that the numerical RG method is computationally more efficient than DNI of equation of motion, because the RG transformation effectively coarse-grains uneven parts of the initial data, and drives more quickly all the initial conditions in its basin of attraction towards its fixed point. Another advantage of our numerical RG method is that, because only a small number of time steps and a small size system are needed, there is no difficulty at all in applying it to two or three dimensional systems, whilst if DNI is used, in order to obtain a reasonably accurate solution, a huge number of time steps and a large system is required.

We have also applied this numerical RG to study the universal behavior of the long-time asymptotics. We choose a number of different initial conditions, sufficiently localized and integrable, *i.e.*, with $\int dx\ u(x,0) = Q_0$ bounded, such as a Lorentzian, step function, and Gaussian, or even those initial distributions with a major peak at origin and other minor multiple peaks elsewhere. In all cases, as long as the initial conditions are well-localised, the numerical RG yields the same anomalous exponents and scaling functions for the long-time asymptotic dynamics, although the rates at which they converge are different. It is also found that the long-time asymptotic behavior is independent of specific numerical schemes, explicit or implicit; nevertheless, the explicit ones are less time-consuming than the implicit ones, and for this reason we prefer to adopt the explicit scheme in all our simulations.

## 3. Numerical Method for Travelling Waves

In this section, we present a numerical RG transformation for travelling waves interpolating between stable and unstable states. In earlier work, we applied a variational principle and perturbative RG to study the dynamical velocity selection mechanism,[10] where it was proposed that physically relevant travelling-wave solutions must be structurally stable in a precise sense.

We define a RG transformation $R_{b,v}$ on the space of functions $u(x,t)$ at some time $t$ by $u'(x,t) = R_{b,v}\{u(x,t)\}$, depending on two parameters, a dilation parameter $b$ and a speed $v$. Two steps follow in this transformation:

(1) Evolve the function $u(x,t)$ forward over a finite time, from $t$ to $t' = (b+1)t, b > 0$, using the governing PDE, and call the result $u'(x,t')$.

(2) Rescale $x$ by defining $x' = x + vbt$, *i.e.* shifting $x$ by an amount of displacement $v(t'-t) = vbt$, where $v$ must be varied to search for the fixed point.

Unlike the similarity case, there is no need to rescale the function $u$ itself, because the rescaling factor $Z(b)$ is actually equal to one (This corresponds to the exponent $\alpha = 0$ if we transform the travelling-wave form $f(x-vt)$ into the similarity form $T^{-\alpha} f(XT^{-v})$ by $T = \log t, X = \log x$.) Thus, we have

$$u'(x,t) = u(x + vbt, (b+1)t). \tag{3.1}$$

This transformation is useful in that any fixed point of $R_{b,v}$ is a travelling-wave solution. This RG transformation generates a semi-group ($b > 0$): $R_{b_1,v} R_{b_2,v} = R_{b_1+b_2,v}$. A fixed point will be reached, after an infinite number of RG iterations, if the initial conditions are in the basin of attraction of it, *i.e.,*

$$u^*(x,t) = u\left(x + vnbt, (nb+1)t\right), \qquad n \to \infty. \tag{3.2}$$

By translationally shifting the solution by an amount $v(nb+1)t$, we have $u^*(x,t) = u(x - vt)$, as $n \to \infty$. As we iterate this RG transformation, we must vary the unknown velocity $v$ to find the fixed point by requiring that $u(0,t) = u(d,t')$, where $t' = (1+b)t$, $d = v_0(t'-t) = v_0 bt$, and $v_0$ is the desired velocity. Only when the profile is shifted back along the $-x$ direction by an appropriate displacement will the shifted solution coincide with the original one. If we were to choose a velocity $v$ larger (or smaller) than $v_0$, then the solution would be overshifted (or undershifted) back and $u(vbt,t')$ would be smaller (or greater) than $u(0,t)$. We emphasize that this is completely different from DNI, in that during RG iterations the unknown parameter $v$ is being varied and we are taking the coarse-grained data as our new initial conditions.

The numerical discretization procedure is almost identical with that for the similarity case. Due to discretization, the number of grid sites by which we shift back to rescale $x$ must be an integer $l$, and we have to take a varying mesh size at each RG iteration, $\Delta x = v/m$ with $m$ a large integer number, so that $l = vt_0/\Delta x = mt_0$ is an integer. Typically we choose $5 \leq l \leq 20$ to guarantee a sufficiently high accuracy of computation. Just as in the similarity case, given an estimated interval for $v$, $v_1 < v < v_2$, a root-finding algorithm is used to pick out the dynamically selected velocity $v^*$, by regarding the difference of $u(x=0,t_0)$ in the last two RG iterations as a function of $v$, $\Delta u(0,t_0) = f(v)$, and requiring $f(v^*) = 0$.

To illustrate how this discrete numerical RG transformation scheme works, we present Fisher's generalized population model,[19]

$$\partial_t u = \partial_x^2 u + u(1-u)(1+\nu u), -1 \le \nu \le +\infty, \tag{3.3}$$

where $u = 0, 1$ are two steady-state points.

When $\nu = -1$, this equation reduces to the well-known Fisher-Kolmogorov-Petrovsky-Piskunov equation,[20] for which rigorous results are known.[21] If the initial condition $u(x, t = 0)$ is assumed to be well-localized and decays as fast as $e^{-qx}$ for large $x$, then for $q \ge 1$, the front velocity asymptotically approaches the value $v = 2$, while for $q < 1$, the asymptotic speed is $v = q + 1/q > 2$. In figure 2, we show numerical RG calculations for the velocity $v$, compared with the exact analytical result above. In doing so, we take $\Delta t, \Delta x$ even as large as $0.05, 1.0$, respectively, and use only a small number of grid points to discretize $t$ and $x$. Our simulation shows that the numerical RG transformation not only drives the solution towards the fixed point–travelling-wave solution more quickly than DNI, but also uniquely determines the correct velocity $v$. The typical relative error is only about 0.5% in this case.

Now we move on to the general case with $-1 \le \nu \le +\infty$. It is well established that in this case, there exists a transition, as the parameter $\nu$ crosses the transition point $\nu_c = 2$, and the corresponding velocity is $v = 2$ for $-1 \le \nu \le 2$, and $v = (2+\nu)/\sqrt{2\nu}$ for $\nu \ge 2$. Starting with sufficiently localized initial conditions ( e.g. a standard step function $u(x, t = 0) = \Theta(-x)$ ), we have performed the numerical RG simulation, and plot our results in figure 3, in comparison with the exact result. It demonstrates that our numerical RG transformation does predict a transition between linear-marginal-stability (pulled) and nonlinear-marginal-stability (pushed) cases and uniquely selects the correct front velocity for most natural initial conditions.

LYC would like to thank Tong Zheng and Qing Zhou for help on numerical codes. The authors gratefully acknowledge support from the National Science Foundation through grant number NSF-DMR-89-20538, administered through the Illinois Materials Research Laboratory. LYC also acknowledges the support of the Institute For Theoretical Physics through the NSF Grant No. PHY89-04035 during the final stages of the preparation of the manuscript.

# FIGURE CAPTIONS

Fig. 1. The anomalous scaling exponent $\alpha$ as a function of the number of numerical RG iterations $N$, in the case of $\epsilon = 1.0$ for the Barenblatt equation. Shown is the convergence to the exact value $\alpha = 0.6975$ for three different initial conditions with comparable width: Lorentzian (filled circles), Gaussian (filled squares) and top hat (filled triangles).

Fig. 2. The propagation velocity $v$ as a function of decay rate $q$ of initial conditions. The points determined by the numerical RG method are denoted by $\bullet$. The continuous curve is the exact result.

Fig. 3. The propagation velocity $v$ plotted as a function of $\nu$. The full curve represents the exact result, whilst data points determined by our numerical RG are denoted by $\bullet$.

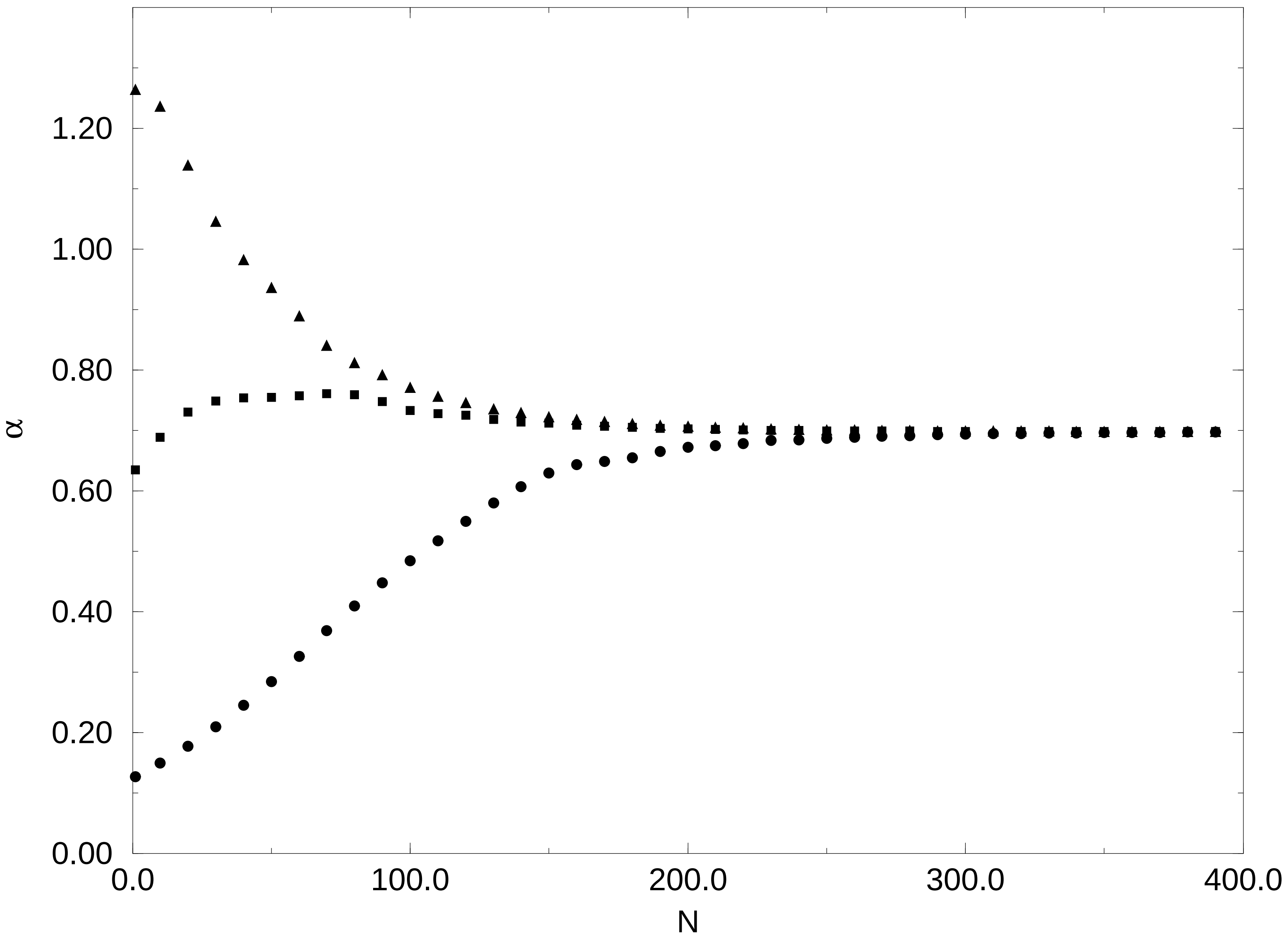

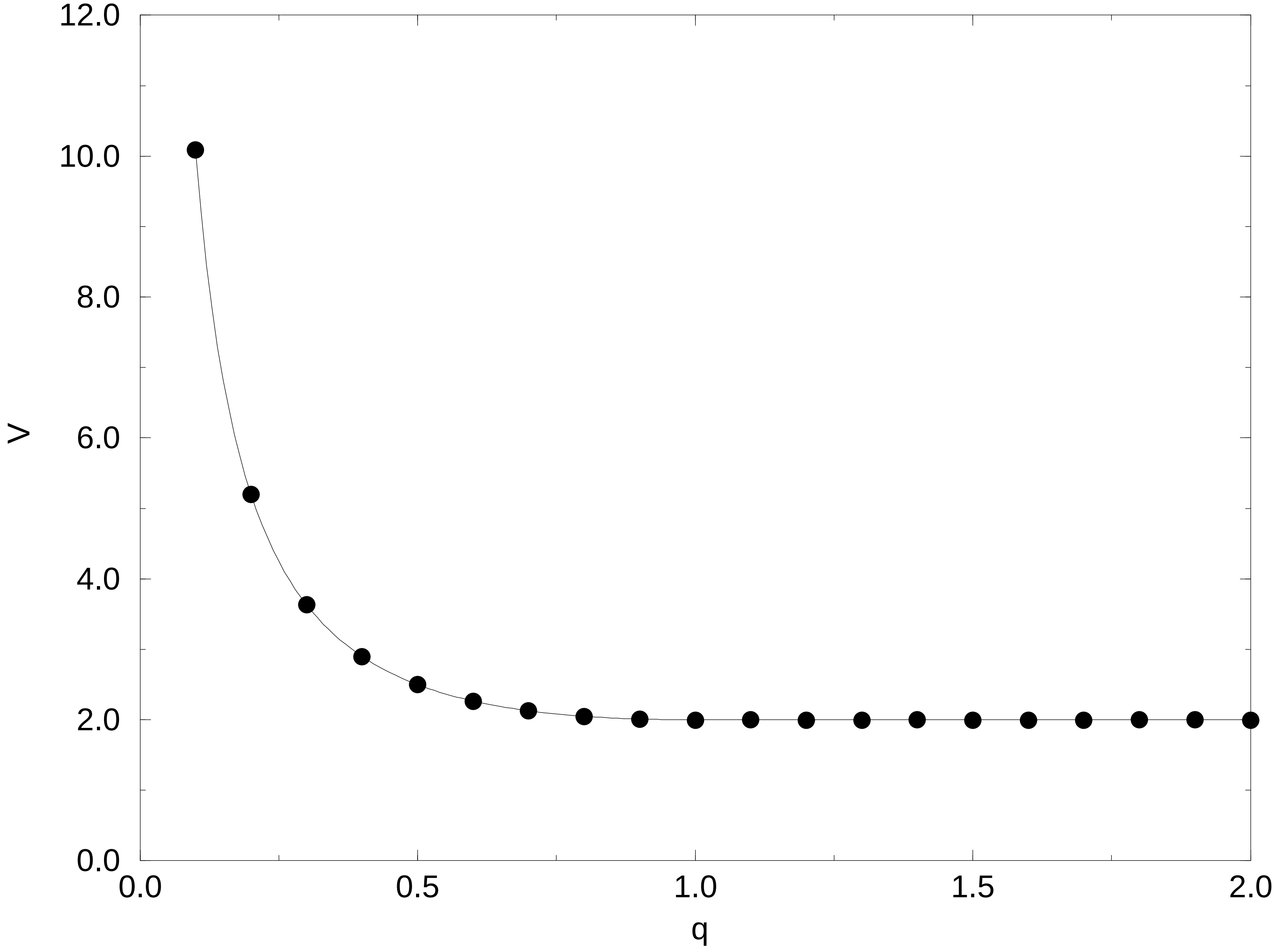

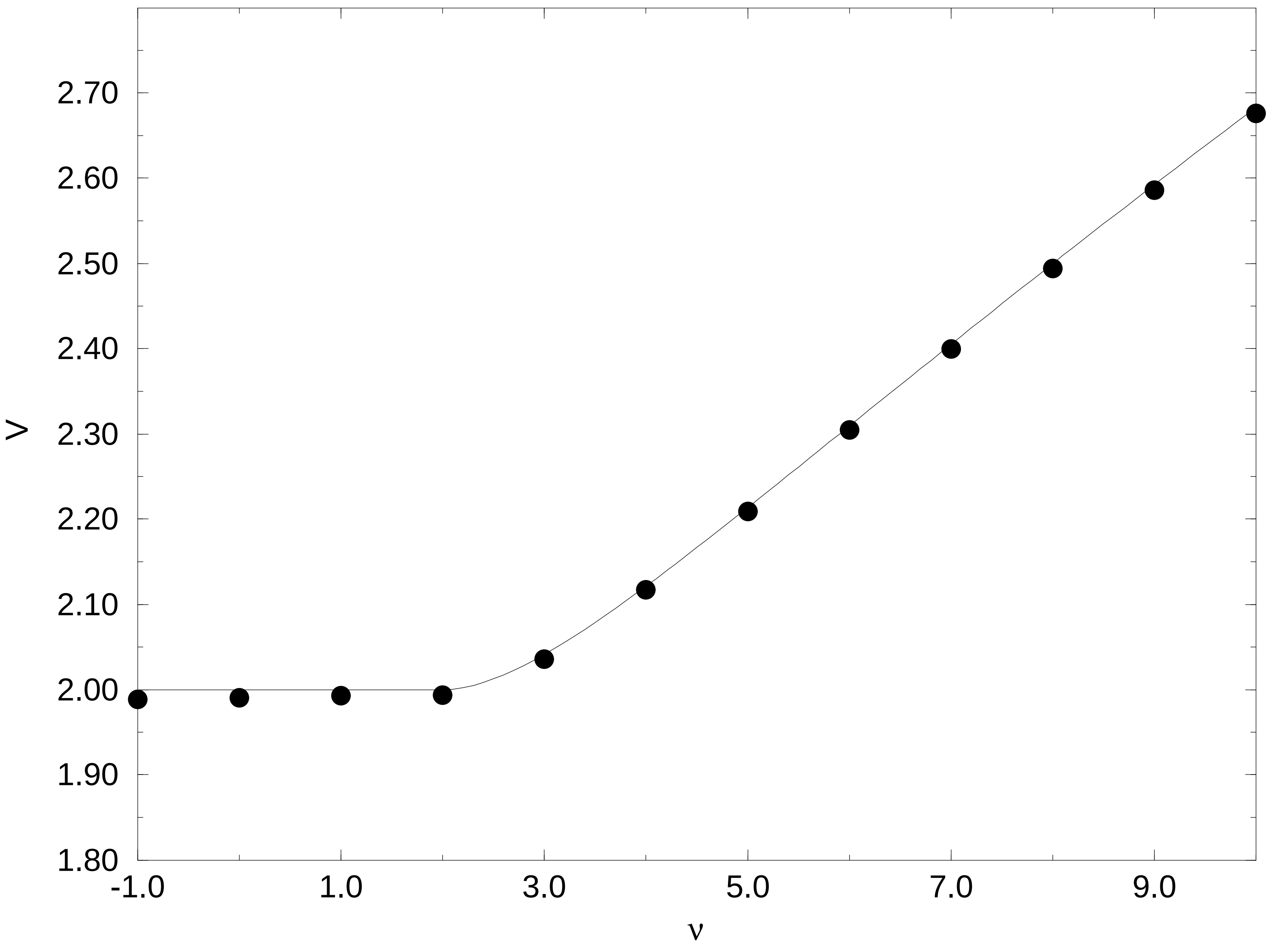